# A CMOS Probabilistic Computing Chip With in-situ Hardware Aware Learning

Jinesh Jhonsa, William Whitehead, David McCarthy, Shuvro Chowdhury, Kerem Camsari, Luke Theogarajan

University of California Santa Barbara, Santa Barbara, USA

**Abstract**

This paper demonstrates a probabilistic bit physics-inspired solver with 440 spins configured in a Chimera graph and occupying an area of 0.44 mm². Area efficiency was maximized through a current-mode implementation of neuron update circuit, standard cell design for analog blocks pitch-matched to digital block, and a shared power supply for digital and analog components. Process variation related mismatches introduced by this approach were effectively mitigated using a hardware-aware contrastive divergence algorithm during training. We validate the chip's ability to perform probabilistic computing tasks, such as modeling logic gates and full adders and optimization tasks, such as Max-Cut. demonstrate its potential for AI and machine learning.

## Introduction

Probabilistic bits (p-bits) have emerged as a hardware friendly approach for solving optimization problems, machine learning, quantum inspired computing and AI [1]. Ideally the probabilistic computation is performed by a stochastic device [2]. However, device integration can be cumbersome, though some implementations overcome some of these issues [5]. In this work, we present an all-CMOS implementation of probabilistic computing. While there has been some excellent work on spin-based circuits for solving certain optimization problems, spins evolve deterministically, and temperature is modeled using randomness for simulated annealing [10]. To the best of our knowledge this is the first fully CMOS p-bit based chip. Large-scale implementation of 8-bit precision p-bit for on-chip probabilistic computing is constrained by neuron update circuit requiring low-area implementations to enable packing 440 p-bits in 0.44 mm². Each Pbit computes two fundamental equations,

$$I_i = \sum_{i \neq j} J_{IJ} m_j + h_i m_i \qquad (1)$$

$$m_i = \text{sgn}(\tanh \beta I_i + Rand(+1, -1)) \qquad (2)$$

We utilize a hardware friendly Chimera graph topology as shown in fig.1 pioneered by D-Wave rather than the interconnect intensive all-to-all topology. The Chimera topology enables embedding many graphs in contrast to the more commonly used King's graph. Each unit cell of the Chimera graph is a 4:4 restricted Boltzmann machine (RBM). Each RBM cell also receives inputs from neighbors. Our design philosophy was driven by two major considerations, 1) area and 2) automation. The first constraint was met by utilizing current mode implementation of eqns. 1 & 2. The second constraint was met by adopting a standard cell design for all analog blocks, pitch matched to the digital blocks. The analog blocks are then treated like digital blocks by the automated place and route vastly simplifying the layout (see Fig. 2b). The use of this approach necessitates sharing the power supply between the analog and digital block for area efficiency. One downside to the approach is the analog circuits are not matched. While this design methodology may not be suitable for general mixed-signal design, it lends itself to the probabilistic computing environment.

The 440 spins are arranged in a 7x8 array of Chimera unit cells, with one cell substituted by bias circuits and SPI interfaces for loading weights and reading spin values. Digital weights (8 bits) were converted to analog bias values using a MOS transistor-based R-2R digital to analog converter (DAC) (see Fig. 3). The MOS R-2R DAC was chosen due to its high area -efficiency and low complexity. However, it should be noted that the choice of using a low supply voltage (1V) and lack of any circuit techniques to improve output resistance will lead to some mismatch issues. Since the topology is an undirected graph, to save area the current was converted into a bias voltage and distributed to the respective nodes. Since setting the weight to zero might not necessarily remove a connection due to mismatch, an enable bit was added. The scale for coupling weights, bias weight, random number and tangent hyperbolic are independently set using external resistors (see bias generator in Fig 6). Implementing equation 1 requires the multiplication of the weights by the spin value. Since the weight is represented by a current, a current mode Gilbert multiplier can be used (see multiplier in Fig. 5), resulting in a differential representation. The differential format enables easy representation of bipolar weights. Summation results directly from current summation by connecting the output nodes together. Each node has 6 current inputs summed on the output node (see Fig. 1). In addition to this there is a bias current input to implement (2), the bias current branch utilizes the same current DAC circuit as the coupling term. For the p-bit representation the summed current needs to serve as the input to a tangent hyperbolic (tanh) function. We used a modified fully differential winner take-all circuit, see Fig. 4 [3], to both pin the voltage enabling better current matching and perform a tanh transformation of the summed current. Each branch implements a Fermi function of the difference of currents and a subtraction then yields the required tanh function. Before performing the subtraction, a differential random current with a uniform distribution was summed to the output node, generated by a DAC circuit identical to the one utilized for the weight/bias, see Fig. 5. The value of the RNG DACs are set by decimated linear feedback shift registers (LFSR) [4], which generate a new pseudorandom value in every bit position every clock cycle. Bitstreams from two LFSRs clocked at 200 MHz were used as 64 unique random clocks of which 55 were used to drive a 32 bit LFSR in each unit cell of the Chimera graph. Since each 32-bit LFSR yields only 4 unique 8-bit random numbers, the vertical nodes of the chimera graph received a normal bit sequence while the horizontal nodes received a reversed bit sequence. While there is a possibility the reverse sequence and the original sequence could have some correlation, there was no noticeable degradation in chip performance using this method. The summed current from the tanh circuit and the random number generator was fed into a winner take-all based comparator and then further amplified using a self-biased fully differential comparator (see Fig. 6).

We used hardware aware learning based on contrastive divergence [1]. The algorithm is pictorially represented in Fig. 7a. The measured learning of a AND gate is shown in Fig. 7b,c. The average value of the spins should produce a tanh function when the bias is swept. We utilized this to calculate the mismatch on-chip, as shown in Fig. 8a. The probability distribution of a full adder was also implemented to demonstrate the chip's ability to perform complex computations. All 440-spins were then utilized in a simulated annealing experiment of a Sherrington-Kirkpatrick spin-glass. Figure 9a show the energy of the function decreasing as the annealing proceeds. Annealing temperature was controlled by a voltage ($V_{temp}$). We also performed a Max-Cut problem as shown in Fig. 9b. A comparison table (Table 1) highlights the features of the chip. A chip micrograph is shown in Fig. 10. machine learning applications

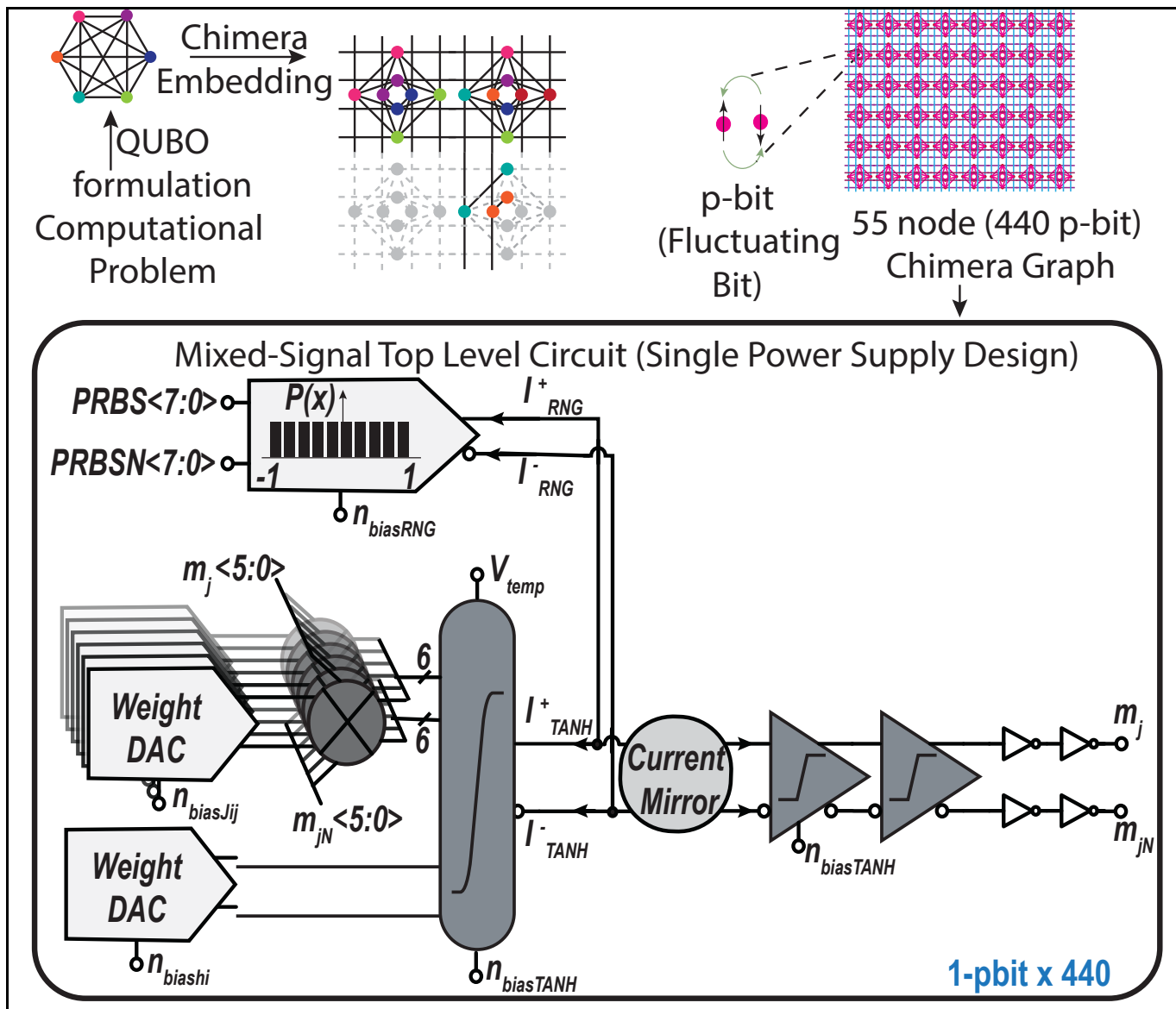

Fig. 1. Schematic of a single probabilistic compute unit. Also shown is the mapping of an optimization problem into a Chimera graph.

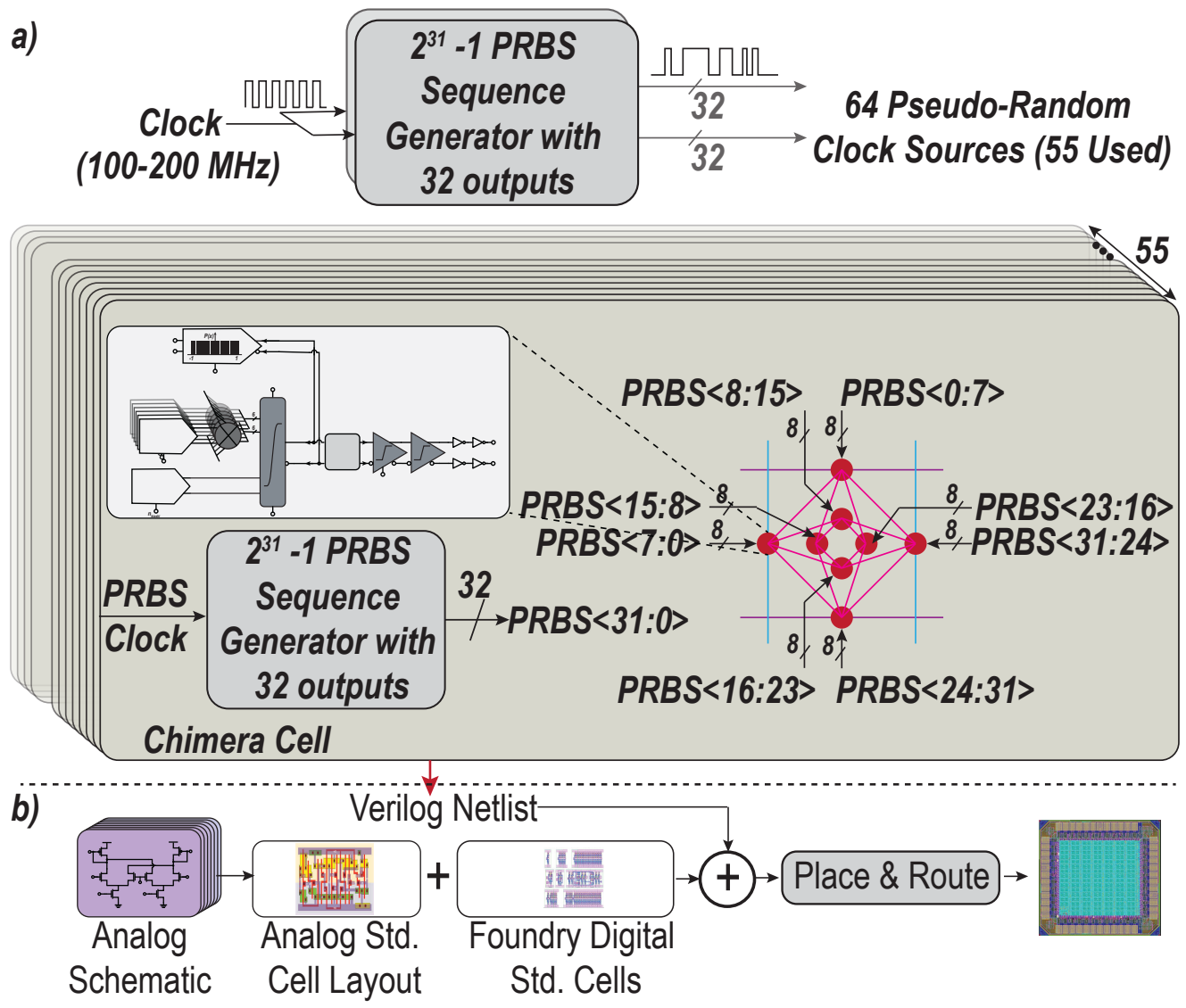

Fig. 2. a) Full chip architecture b) Design automation methodology used

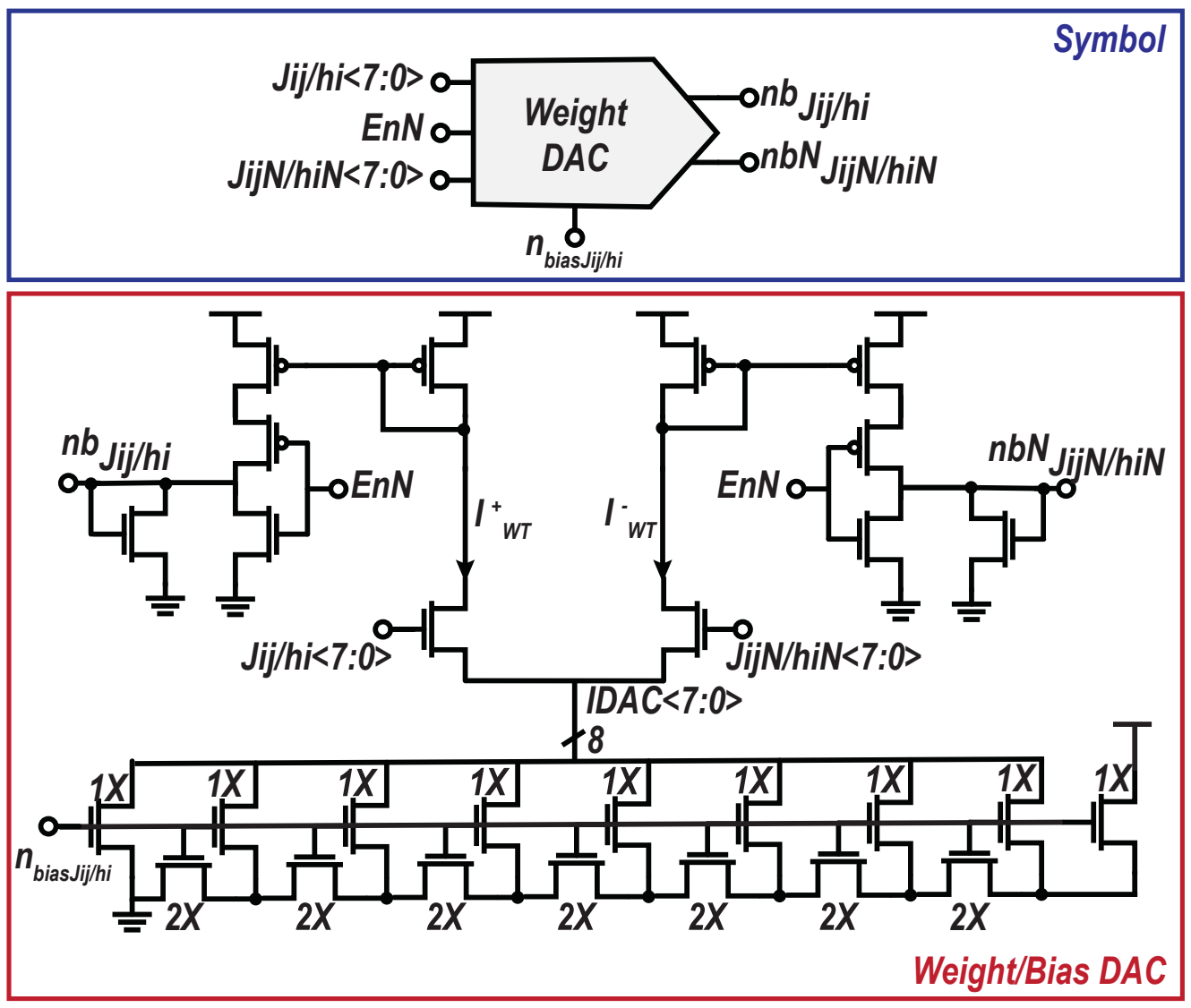

Fig. 3. Schematic of the Weight/Bias DAC

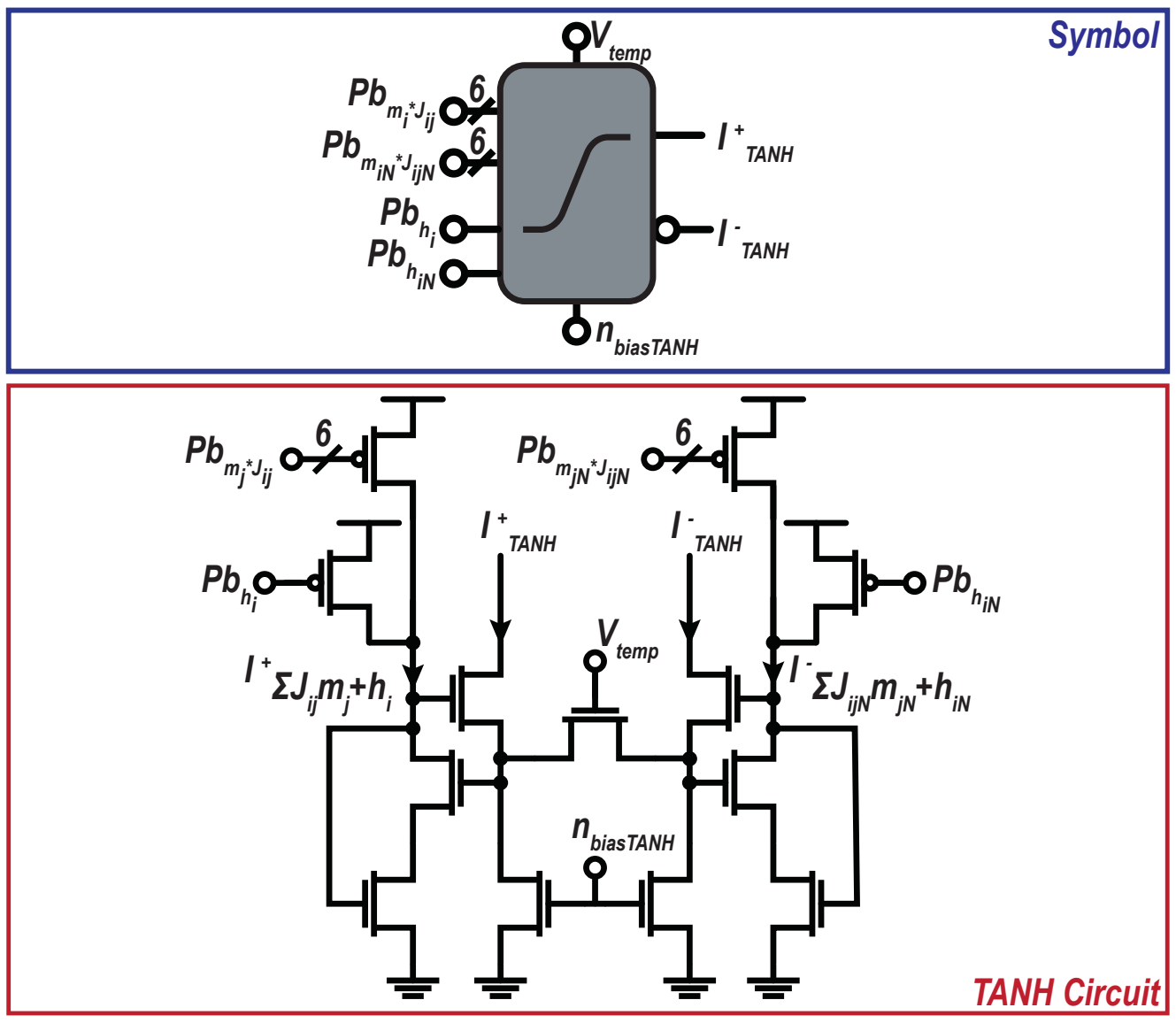

Fig. 4. Circuit implementation of the tangent hyperbolic.

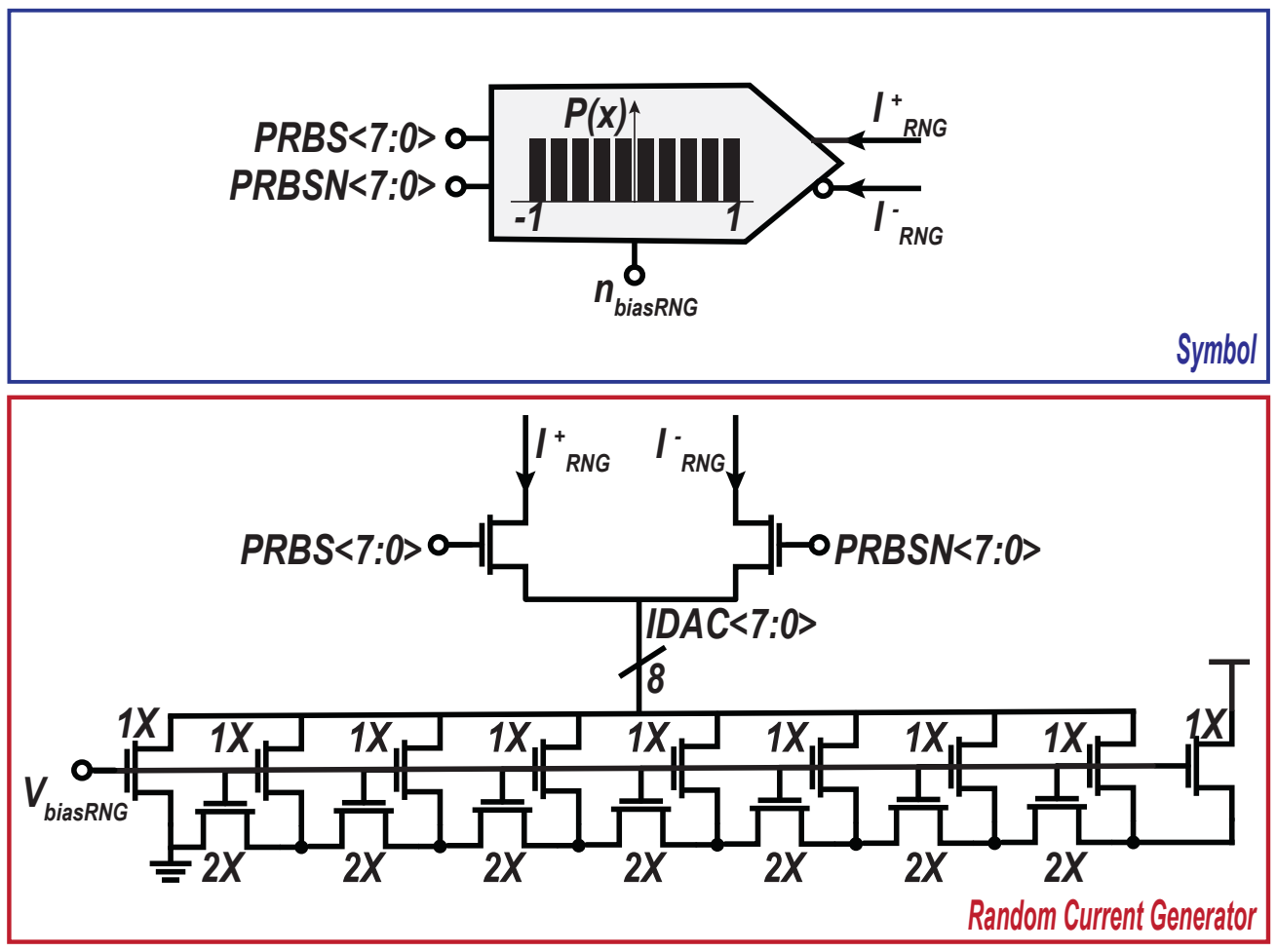

Fig. 5. Implementation of randomness on chip. Random currents are generated using a PRBS and a current-mode DAC..

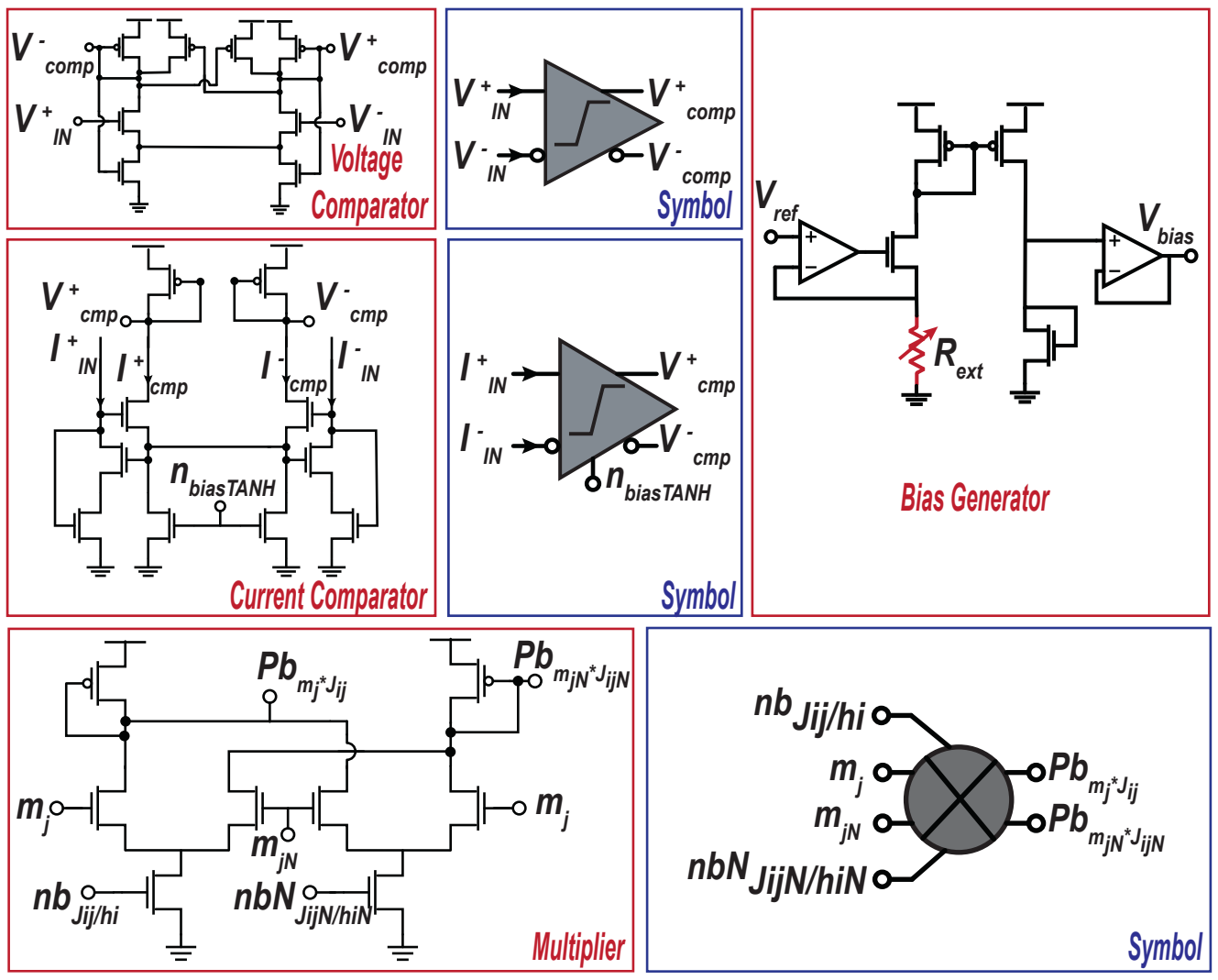

Fig. 6. Schematic of Comparators and Multiplier.

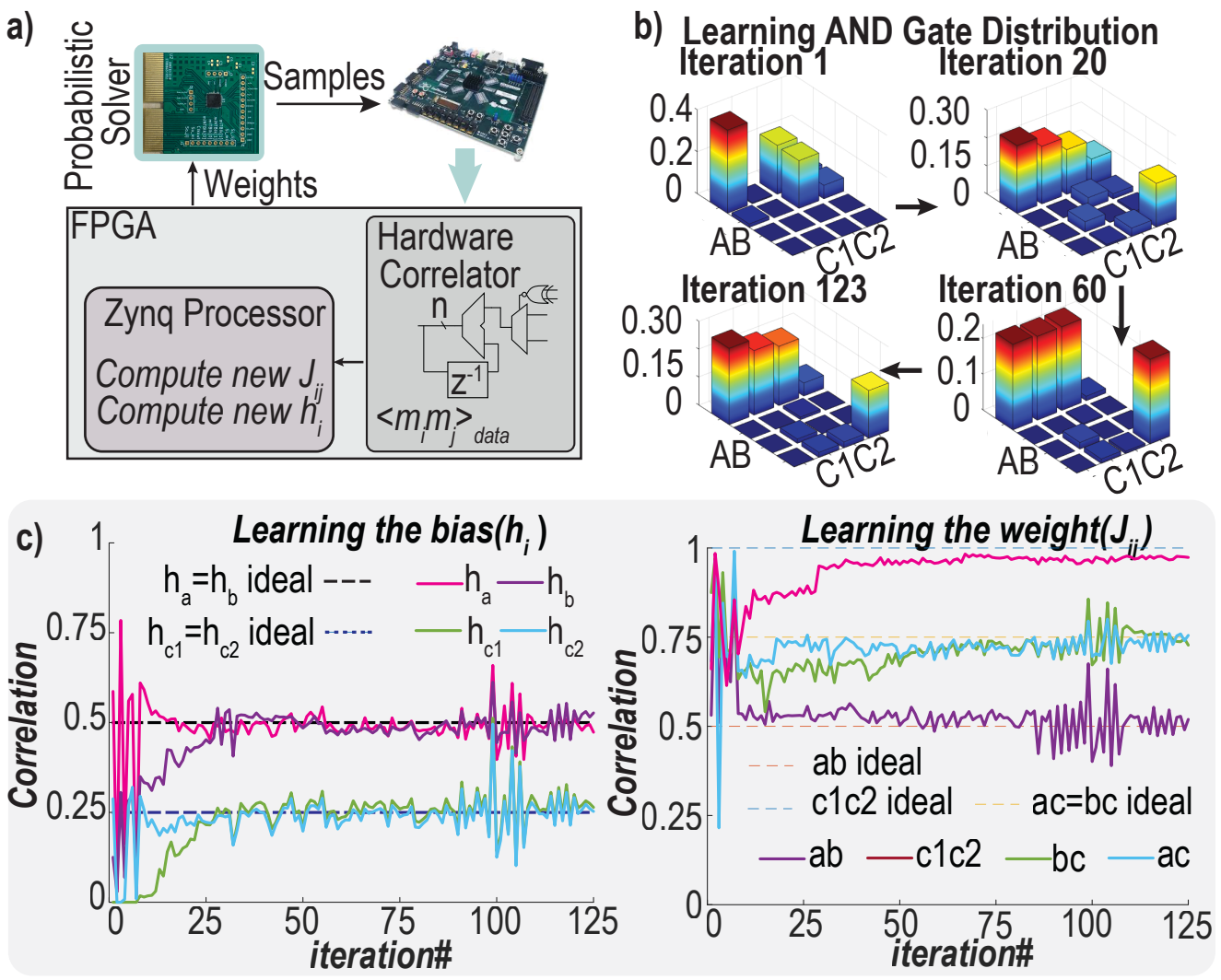

Fig. 7. a) Pictorial representation of the contrastive divergence algorithm utilized for learning. b) Measured AND gate distribution as learning proceeds. c) Correlation convergence during learning

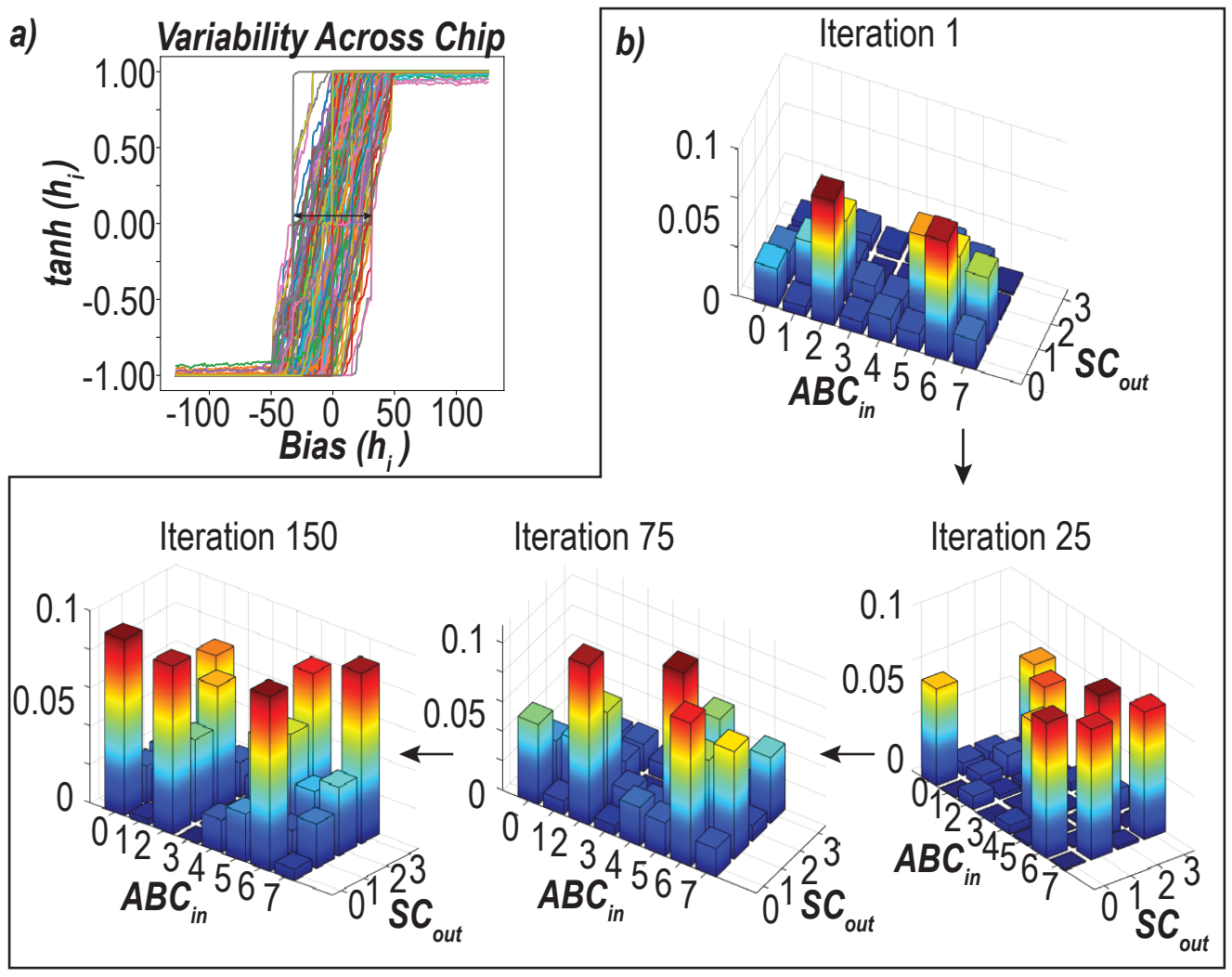

Fig. 8. a) Variability across the chip measured by taking the average of samples as function of bias. b) Measured Adder probability distribution as learning proceeds.

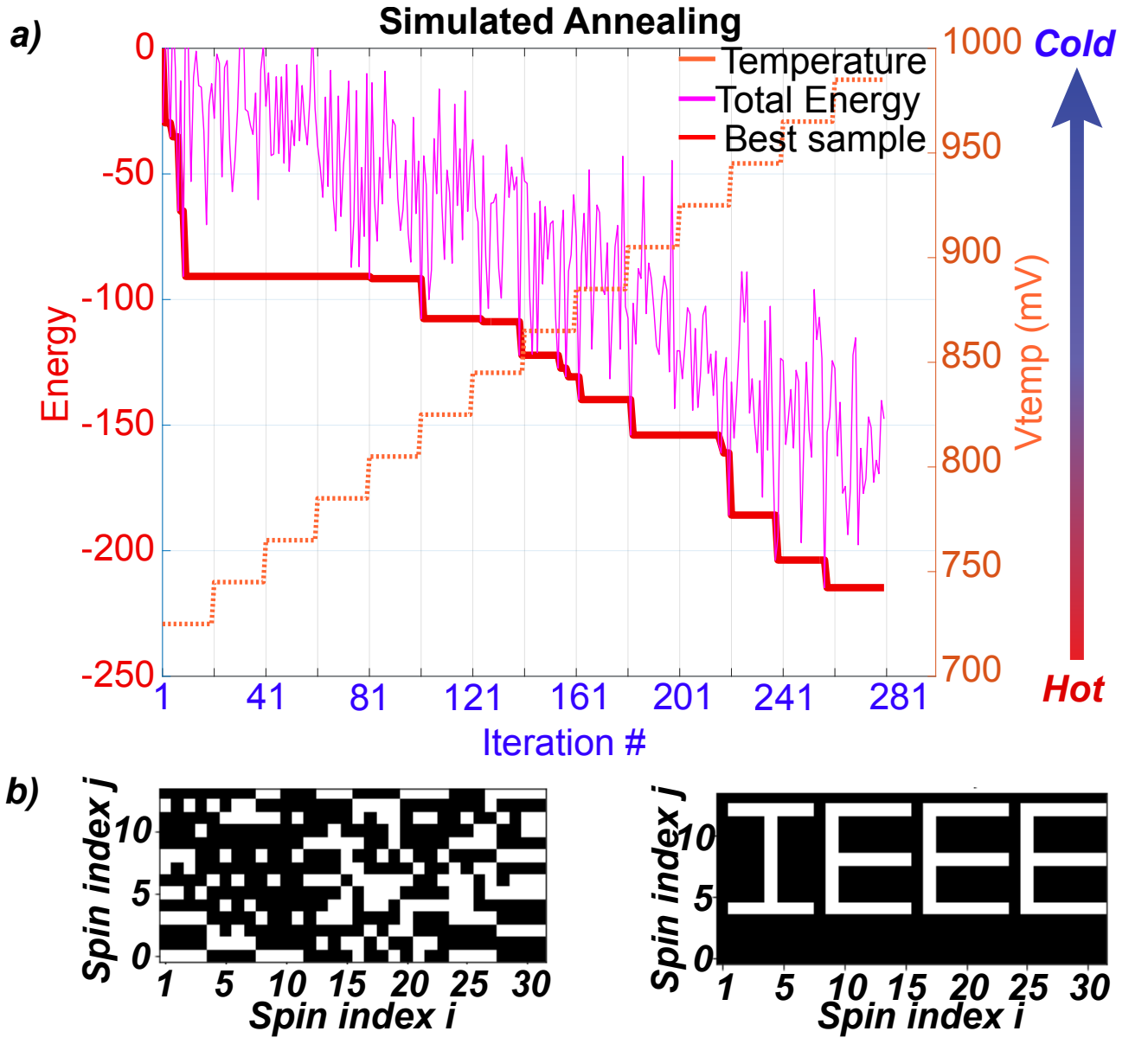

Fig. 9. a) Simulated Annealing b) Max-Cut

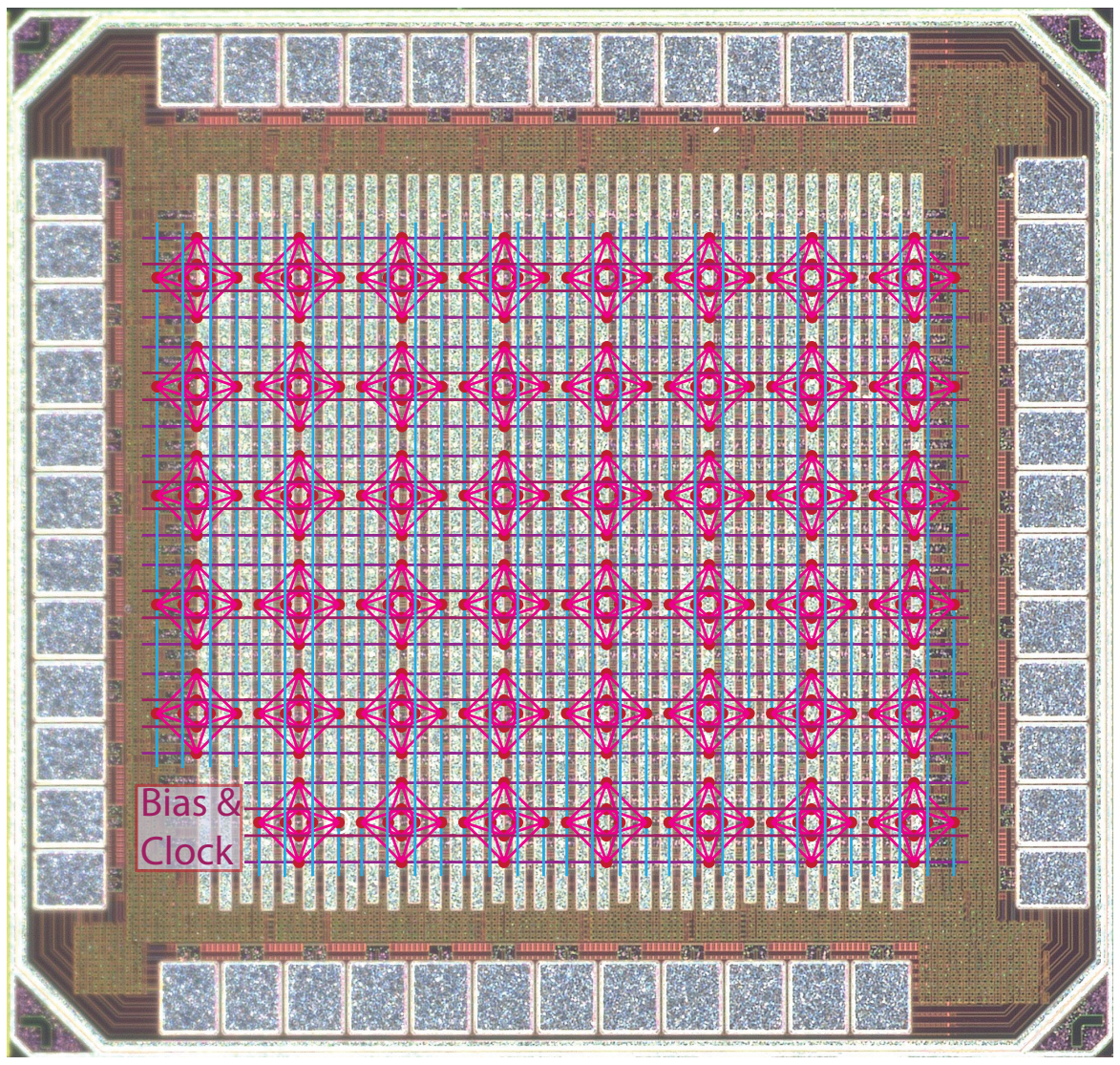

Fig. 10. Chip microphotograph

| | VLSI 20 [6] | ISSCC 23[7] | JSSC 22 [8] | ISSSC 24[9] | This Work |
|---|---|---|---|---|---|
| Technology (Circuit Type) | 65nm (Mixed-Signal) | 65nm (Mixed-Signal) | 65nm (Mixed-Signal) | 65nm (Mixed-Signal) | 65nm (Mixed-Signal) |
| Spin memory | Ring-Oscillator | CMOS Latch | eDRAM Cell | SRAM Cell | Flip - Flop |
| Spin State update | Analog (ROSC Phase) | Analog (Latch Voltage) | Digital (Binary State) | Analog (Latch Voltage) | Digital (Binary State) |
| Graph Topology | Hexagonal (6x Spins) | Lattice (4x Spins) | King's (8x Spins) | e-Chimera (11xSpins) | Chimera (8x spins) |
| Ising Hamiltonian | No | Latch Equalized | Simulated Annealing | Latch Equalize | Gibbs Sampling |
| Supply | 1V | 0.7-1.05V | 0.9-1.2V | 0.8-1.4V | 1V |
| Spins# (Core size) | 560 0.53mm$^2$ | 1,440 0.44mm$^2$ | 6,400 0.71mm$^2$ | 1,536 0.16mm$^2$ | 440 0.44mm$^2$ |
| TTS | 1-10us | <100ns | 0.05ms | <100ns | 50ns |

TABLE 1. Comparison with state-of-the-art

**Acknowledgements:** K.Y.C. and S.C. have been supported by an ONR-MURI grant N000142312708. JJ, DM, WW and LT would like to acknowledge support from the DARPA PIPES DARPA PIPES program under contract HR0011-19-C-0083. The views, opinions and/or findings expressed are those of the author and should not be interpreted as representing the official views or policies of the Department of Defense or the U.S. Government. Released under Distribution Statement "A"

(Approved for Public Release, Distribution Unlimited).